\documentclass[axioms,article,accept,pdftex,moreauthors]{Definitions/mdpi}

\usepackage[markup=underlined]{changes}

\firstpage{1} 
\pubvolume{13}
\issuenum{3}
\articlenumber{187}
\pubyear{2024}
\copyrightyear{2024}
\externaleditor{Academic Editor: Mariam Zomorodi}
\datereceived{27 January 2024 } 
\daterevised{7 March 2024 } 
\dateaccepted{8 March 2024 } 
\datepublished{13 March 2024} 
\hreflink{https://doi.org/10.3390/\linebreak axioms13030187} 

\usepackage{amsmath}
\usepackage{natbib}
\usepackage{braket}

\usepackage{tikz}
\usetikzlibrary{arrows.meta,
                chains,
                positioning,
                shapes.geometric
                }

\usepackage[utf8]{inputenc} 
\usepackage[T1]{fontenc}    
\usepackage{url}            
\usepackage{booktabs}       
\usepackage{amsfonts}       
\usepackage{nicefrac}       
\usepackage{tabularx}
\usepackage{microtype}      
\usepackage{xcolor}         
\usepackage{graphicx}
\usepackage{subcaption}
\usepackage{float}

\Title{Hybrid Quantum Vision Transformers for Event Classification in High Energy Physics}

\TitleCitation{Hybrid Quantum Vision Transformers for Event Classification in High Energy Physics}

\Author{
Eyup B. Unlu $^{1,}$*\orcidF{},  
Marçal Comajoan Cara $^{2}$\orcidB{},
Gopal Ramesh Dahale $^{3}$\orcidC{},
Zhongtian Dong $^{4}$\orcidG{},\linebreak
Roy T. Forestano $^{1}$\orcidA{},
Sergei Gleyzer $^{5}$\orcidH{},
Daniel Justice $^{6}$\orcidI{},
Kyoungchul Kong $^{4}$\orcidJ{},
Tom Magorsch $^{7}$\orcidK{},
\mbox{Konstantin T. Matchev} $^{1}$ \orcidD{} 
and
Katia Matcheva $^{1}$\orcidE{} 
 }

\AuthorNames{Eyup B. Unlu, Marçal Comajoan Cara, Gopal Ramesh Dahale, Zhongtian Dong, Roy T. Forestano, Sergei Gleyzer, Daniel Justice, Kyoungchul Kong, Tom Magorsch, Konstantin T. Matchev and Katia Matcheva}

\AuthorCitation{Unlu, E.B.; Comajoan Cara, M.; Dahale, G.R.; Dong, Z.; Forestano, R.T.; Gleyzer, S.; Justice, D.; Kong, K.; Magorsch, T.; Matchev, K.T.; et al.}

\address{$^{1}$ \quad Institute for Fundamental Theory, Physics Department, University of Florida, Gainesville, FL 32611, USA; roy.forestano@ufl.edu (R.T.F.); matchev@ufl.edu (K.T.M.); matcheva@ufl.edu~(K.M.) \\
$^{2}$ \quad Department of Signal Theory and Communications, Polytechnic University of Catalonia,\linebreak  08034 Barcelona, Spain; marcal.comajoan@estudiantat.upc.edu
\\
$^{3}$ \quad Indian Institute of Technology Bhilai, Bhilai  491001, Chhattisgarh, India; gopald@iitbhilai.ac.in\\
$^{4}$ \quad Department of Physics \& Astronomy, University of Kansas, Lawrence, KS 66045, USA; cdong@ku.edu~(Z.D.); kckong@ku.edu~(K.K.)\\
$^{5}$ \quad Department of Physics \& Astronomy, University of Alabama, Tuscaloosa, AL 35487, USA; sgleyzer@ua.edu\\
$^{6}$ \quad Software Engineering Institute, Carnegie Mellon University, 4500 Fifth Avenue, Pittsburgh, PA 15213, USA; dljustice@sei.cmu.edu \\
$^{7}$ \quad Physik-Department, Technische University of München, James-Franck-Str. 1, 85748 Garching, Germany; tom.magorsch@tum.de
}

\corres{Correspondence: eyup.unlu@ufl.edu}

\abstract{Models based on vision transformer architectures are considered state-of-the-art when it comes to image classification tasks. However, they require extensive computational resources both for training and deployment. The problem is exacerbated as the amount and complexity of the data increases. Quantum-based vision transformer models could potentially alleviate this issue by reducing the training and operating time while maintaining the same predictive power. Although current quantum computers are not yet able to perform high-dimensional tasks, they do offer one of the most efficient solutions for the future. In this work, we construct several variations of a quantum hybrid vision transformer for a classification problem in high-energy physics (distinguishing photons and electrons in the electromagnetic calorimeter). We test them against classical vision transformer architectures. Our findings indicate that the hybrid models can achieve comparable performance to their classical analogs with a similar number of parameters.}


\begin{document}

\section{Introduction}
The first transformer architecture was introduced in 2017 by Vaswani {et al.} in a famous paper ``Attention Is All You Need''  \cite{all-you-need}. The new model was shown to outperform the existing state-of-the-art models by a significant margin for the English-to-German and English-to-French {\tt newstest2014} tests. Since then, the transformer architecture has been implemented in numerous fields and has become the go-to model for many different applications such as sentiment analysis \cite{sentiment} and question answering \cite{question_answering}.

The {vision transformer} architecture can be considered as the implementation of transformer architecture for image classification. It utilizes the encoder part of the transformer architecture and attaches a multi-layer perceptron (MLP) layer to classify images. This architecture was first introduced by Dosovitskiy {et al.} in the paper ``An Image is Worth 16x16 Words: Transformers for Image Recognition at Scale'' \cite{ViT}. It was shown that in a multitude of datasets, a vision transformer model is capable of outperforming the state-of-the-art model ResNet152x4 while using less computation time to pre-train. Similar to their language counterparts, vision transformers became the state-of-the-art models for a multitude of computer vision problems such as image classification \cite{vt1} and semantic segmentation \cite{vt2}. 

However, these advantages come at a cost. Transformer architectures are known to be computationally expensive to train and operate \cite{transformer1}. Specifically, their demands on computation power and memory increase quadratically with the input length. A number of studies have attempted to approximate self-attention in order to decrease the associated quadratic complexity in memory and computation power \cite{approx1,approx2,approx3,randomattention}. There are also proposed modifications to the architecture which aim to alleviate the quadratic complexity \cite{sub1,bigbird,performer}. A recent review of the different methods for reducing the complexity of transformers can be found in \cite{survey1}. As the amount of data grows, these problems are exacerbated. In the future, it will be necessary to find a substitute architecture that has similar performance but demands fewer resources. 

A {quantum machine learning model} might be one of those substitutes. Although the hardware for quantum computation is still in its infancy, there is a high volume of research that is focused on the algorithms that can be used on this hardware. The main appeal of quantum algorithms is that they are already known to have computational advantages over classical algorithms for a variety of problems. For instance, Shor's algorithm can factorize numbers significantly faster than the best classical methods \cite{shor}. Furthermore, there are studies suggesting that quantum machine learning can lead to computational speedups~\cite{boolean,quantumspeedup}.

In this work, we develop a quantum-classical hybrid vision transformer architecture. We demonstrate our architecture on a problem from experimental high energy physics, which is an ideal testing ground because experimental collider physics data are known to have a significant amount of complexity and computational resources represent a major bottleneck \cite{HEPSoftwareFoundation:2017ggl,HSFPhysicsEventGeneratorWG:2020gxw,Humble:2022klb}. Specifically, we use our model to classify the parent particle in an electromagnetic shower event inside the CMS detector. In addition, we will test the performance of our hybrid architecture by benchmarking it against a classical vision transformer of equivalent architecture.

This paper is structured as follows. In Section \ref{section:data}, we {introduce} and describe the dataset. The model architectures for both the classical and hybrid models are discussed in Section~\ref{section:models}. The model parameters and the training are specified in Sections \ref{section:hyper} and \ref{section:training}, respectively. Finally, in Section \ref{section:results} we {present} our results and discuss {their implications} in Section \ref{section:discussion}. We {consider} the future directions for study in Section~\ref{sec:outlook}.

\section{Dataset and Preprocessing Description}
\label{section:data}

The Compact Muon Solenoid (CMS) is one of the four {main} experiments at the Large Hadron Collider (LHC), {which has been in operation since 2009} at CERN. The CMS detector~{\cite{CMS:2008xjf}} {has been recording} the products from {collisions between beams consisting of protons or Pb ions, at several different center-of-mass energy milestones, up to the current 13.6 TeV} \cite{CMS_HeavyIon}. {Among the various available CMS datasets \cite{CMS_Datasets}, we have chosen to study data from proton--proton collisions at 13.6 TeV.} Among the basic types of objects reconstructed from those collisions are photons and electrons, which leave rather similar signatures in the CMS electromagnetic calorimeter (ECAL) {(see, e.g., Ref.~\cite{Benaglia:2014aqa} and references therein)}. A common task in high-energy physics is to classify the resulting electromagnetic shower in the ECAL as a photon ($\gamma$) or electron ($e^-$). In practice, one also uses information from the {CMS tracking system} \cite{CMS:2014pgm} {
and leverages the fact that an electron leaves a track, while a photon does not}. {However,} for the purposes of our study, we shall limit ourselves to the ECAL only.

\begin{figure}[H]
\includegraphics[width=.8\textwidth]{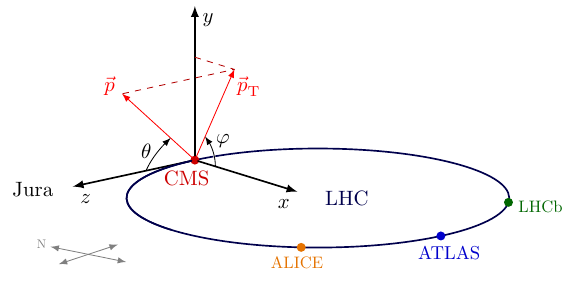}
\caption{{The CMS coordinate system against the backdrop of the LHC, with the location of the four main experiments (CMS, ALICE, ATLAS and LHCb). The $z$ axis points to the Jura mountains, while the $y$-axis points toward the sky. In spherical coordinates, the components of a particle momentum $\vec{p}$ are its magnitude $|\vec{p}|$, the polar angle $\theta$ (measured from the $z$-axis), and the azimuthal angle $\varphi$ (measured from the $x$-axis). The transverse momentum $\vec{p}_T$ is the projection of $\vec{p}$ on the transverse ($xy$) plane. Figure generated with TikZ code adapted from Ref.~\cite{CMS_Coordinate_System}.}}
\label{fig:CMS}
\end{figure}

The dataset used in our study contains the reconstructed hits of 498,000 simulated electromagnetic shower events in the ECAL sub-detector of the CMS experiment {(photon conversions were not simulated)} \cite{Andrews:2018gew}. Half of the events originate from photons, while the remaining half are initiated by electrons. In each case, an event is generated with exactly one particle ($\gamma$ or $e^-$) which is fired from the interaction point with fixed {transverse momentum magnitude $|\vec{p}_T|=50$ GeV, see Figure~\ref{fig:CMS}}. The direction of the momentum $\vec{p}$ is sampled uniformly in {azimuthal angle $-\pi \le \varphi \le \pi$ and pseudorapidity $-1.4\le \eta\le 1.4$, where the latter is defined in terms of the polar angle $\theta$ as $\eta = -\ln \tan (\theta/2)$.}

\vspace{-6pt}

For each event, the dataset includes two image grids, representing energy and timing information, respectively. The first grid gives the peak energy detected by the crystals of the detector in a 32 $\times$ 32 grid centered around the crystal with the maximum energy deposit. The second image grid gives the arrival time when the peak energy was measured in the associated crystal (in our work, we shall only use the first image grid with the energy information.) Each pixel in an image grid corresponds to exactly one ECAL crystal, though not necessarily the same crystal from one event to another. The images were then scaled so that the maximum entry for each event was set to 1.

\begin{figure}[H]
\begin{center}
\includegraphics[width=.245\textwidth]{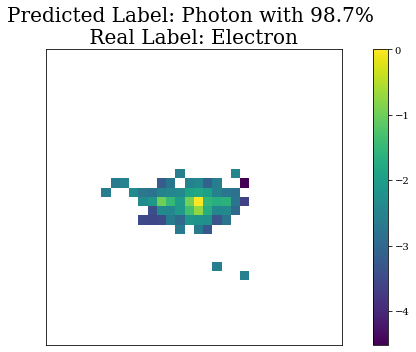}
\includegraphics[width=.245\textwidth]{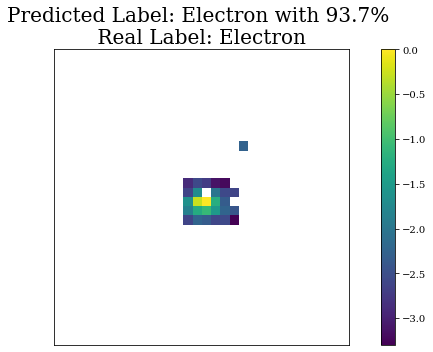}
\includegraphics[width=.245\textwidth]{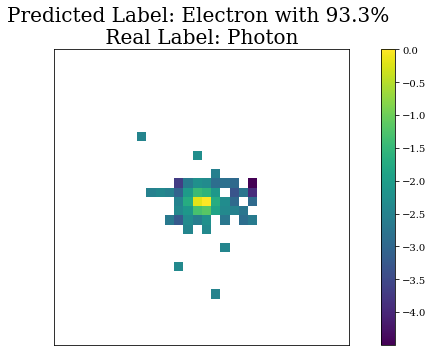}
\includegraphics[width=.245\textwidth]{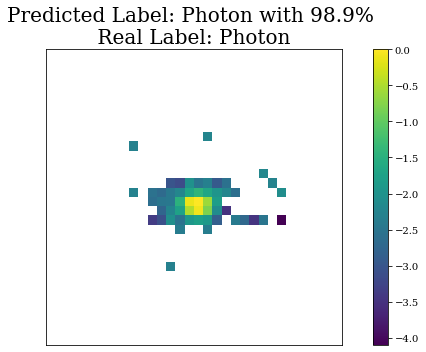}
\\
~~~
\includegraphics[width=.22\textwidth]{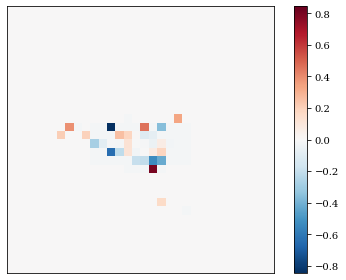}~~~~
\includegraphics[width=.22\textwidth]{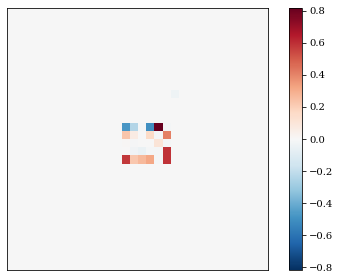}~~~~~
\includegraphics[width=.22\textwidth]{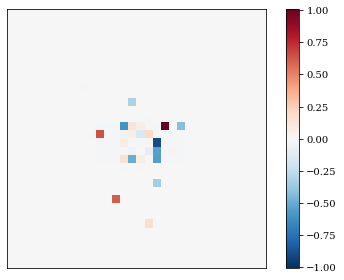}~~~~
\includegraphics[width=.22\textwidth]{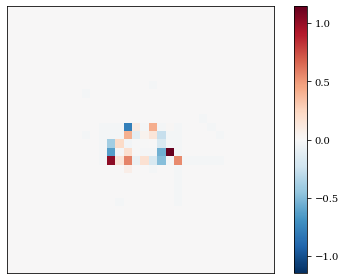}
\end{center}
\caption{Four representative image grid examples from the dataset, in the $(\varphi,\eta)$ plane. The first row shows the image grids for the energy (normalized and displayed in log$_{10}$ scale), while the second row displays the timing information. The titles list the {correct} labels (real electron or real photon), as well as the corresponding labels predicted by one of the benchmark classical models {(see text for more details)}.}
\label{fig:samples}
\end{figure}

Several representative examples of our image data are shown in Figure~\ref{fig:samples}. {The first row shows the image grids for the energy (normalized and displayed in log$_{10}$ scale), while the second row displays the timing information (not used in our study). In each case, the top row in the title lists the label predicted by one of the benchmark classical models, while the bottom row shows the correct label for that instance---whether the image was generated by an actual electron or photon.} 

As can be gleaned from Figure~\ref{fig:samples} with the naked eye, electron--photon discrimination is a challenging task{---for example, the first and third images in Figure~\ref{fig:samples} are wrongly classified.} To first approximation, the $e^-$ and $\gamma$ shower profiles are identical, and mostly concentrated in a 3 $\times$ 3 grid of crystals around the main deposit. However, interactions with the magnetic field of the CMS solenoid ($B=3.8$ T) cause electrons to emit bremsstrahlung radiation, preferentially in $\varphi$. This introduces a higher-order perturbation on the shower shape, causing the {electromagnetic shower profiles \cite{SempereRoldan:2011hpa}} to be more spread out and slightly asymmetric in $\varphi$.

\section{Model Architectures}
\label{section:models}


The following definitions will be used for the rest of the paper and are listed here for~convenience.
\begin{itemize}
\item $n_t$: Number of tokens/patches
\item $d_i$: Flattened patch length
\item $d_t$: Token length
\item $n_{h}$: Number of heads
\item $d_{h}\equiv \frac{d_t}{n_h}$: Data length per head
\item $d_{ff}$: The dimension of the feed-forward network
\end{itemize}

\subsection{General Model Structure}
\begin{figure}[H]
\centering

(a)

\scalebox{0.9}{%
\begin{tikzpicture}[    node distance = -1mm and 3mm,
      start chain = going right,
  circ/.style = {shape=circle,  draw,color=white,fill=orange},
  rect/.style = {rounded corners,draw, align=center,color=white,fill=blue}]
\tikzstyle{every node}=[font=\scriptsize]]

    \begin{scope}[every node/.append style={on chain,thick, join=by -Stealth}]

\node(full){\includegraphics[width=.1\textwidth]{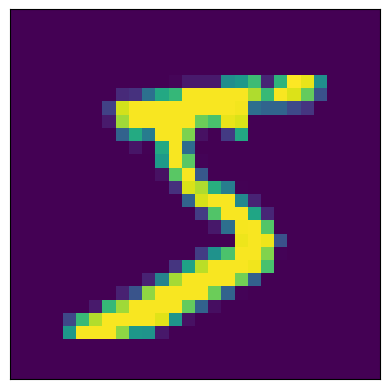}};
\node(patched){\includegraphics[width=.1\textwidth]{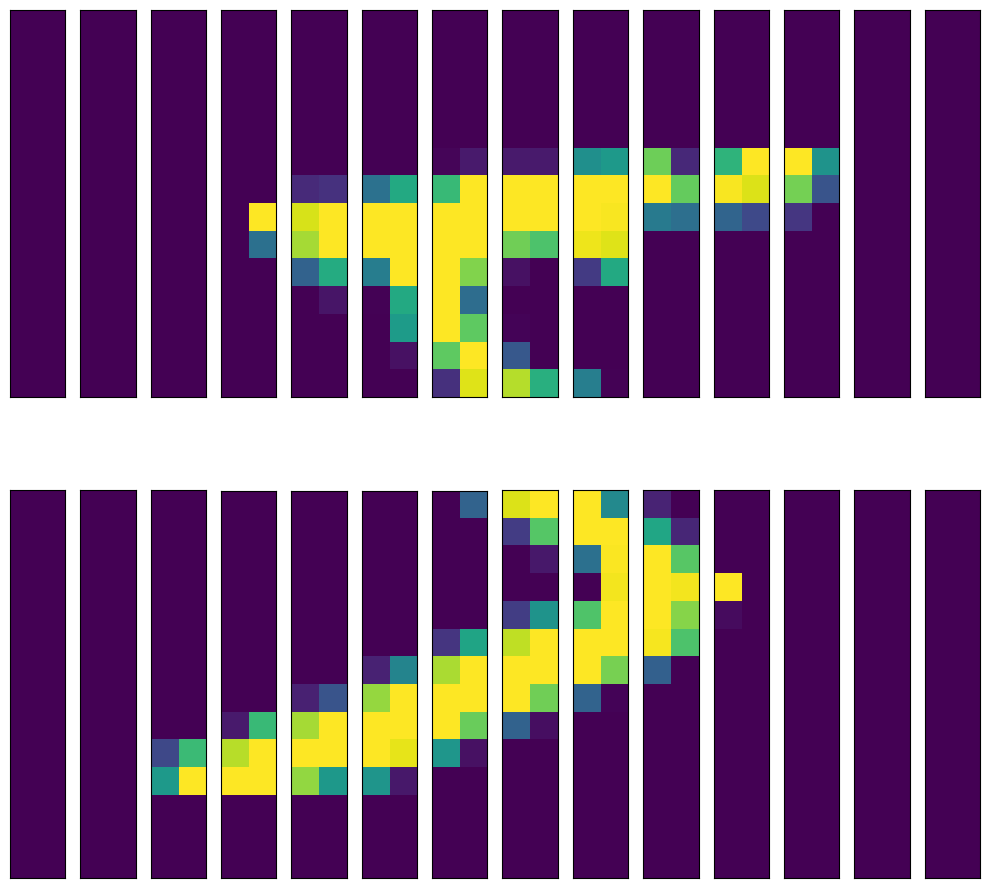}};
\node(flattened){\includegraphics[width=1cm,height=.1\textwidth]{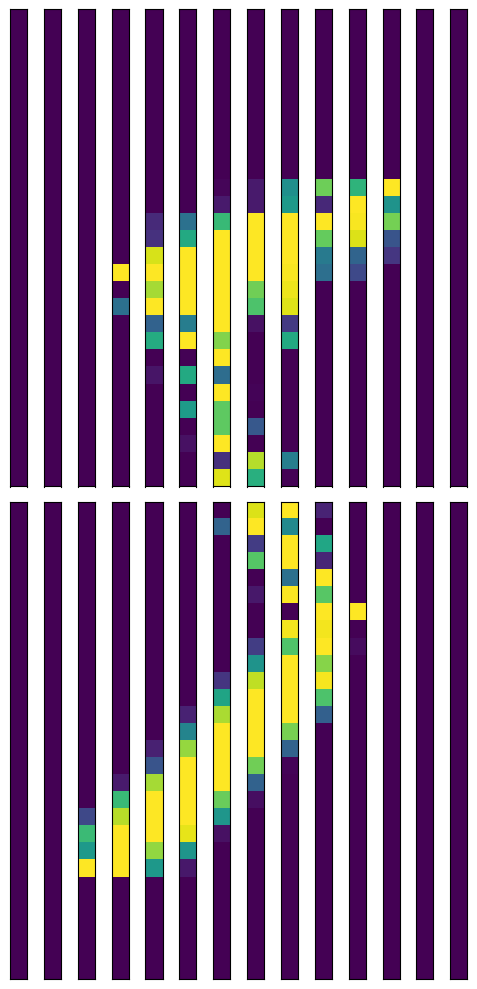}};
\node(concat){\includegraphics[width=1cm,height=.1\textwidth]{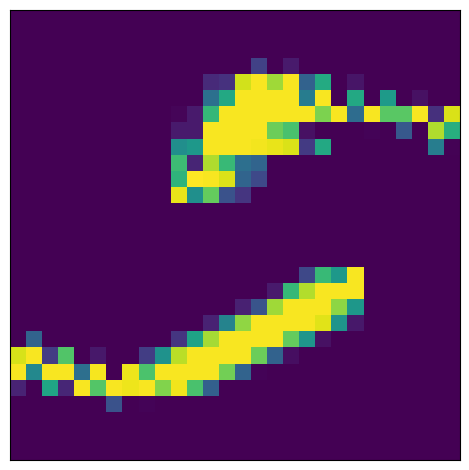}};
\node[rect,align=center] (linear){Linear\\ Embedding};
\node[circ,align=center] (plus){+};
\node[rect,align=center] (enc){Encoder\\ Layer};
\node[rect,align=center] (final){Column-wise Pooling\\ MLP\\(32,LReLU,1)};
    \end{scope}
\node[above= .5cm of plus,align=center] (pos){Positional\\ Embedding \\ \includegraphics[width=.4\textwidth]{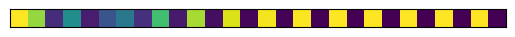}};

\node[rect,below = .75cm of final,fill=orange] (prob){Probabilities};
\node[above left = .15cm of prob.north,thick]{Sigmoid} ;
\node[below = 0cm of enc, align=center]{ x \\ N};
\path[->,thick]
(final) edge (prob) (pos) edge (plus);

\end{tikzpicture}
}

(b)

\scalebox{0.9}{%
\begin{tikzpicture}[    node distance = -1mm and 3mm,
      start chain = going right,
  circ/.style = {shape=circle,  draw,color=white,fill=orange},
  rect/.style = {rounded corners,draw, align=center,color=white,fill=blue}]
\tikzstyle{every node}=[font=\scriptsize]]

    \begin{scope}[every node/.append style={on chain,thick, join=by -Stealth}]

\node(full){\includegraphics[width=.1\textwidth]{Figures/full_image.png}};
\node(patched){\includegraphics[width=.1\textwidth]{Figures/patched.png}};
\node(flattened){\includegraphics[width=1cm,height=.1\textwidth]{Figures/flattened.png}};
\node(concat){\includegraphics[width=1cm,height=.1\textwidth]{Figures/concatenated.png}};
\node[rect,align=center] (linear){Linear\\ Embedding};
\node[rect,align=center] (cat){Concat};
\node[circ,align=center] (plus){+};
\node[rect,align=center] (enc){Encoder\\ Layer};
\node[rect,align=center] (final){Extract Class Token\\ MLP\\(32,LReLU,1)};
    \end{scope}
\node[above= .5cm of plus,align=center] (pos){Positional\\ Embedding \\ \includegraphics[width=.4\textwidth]{Figures/pos_embedding.png}};

\node[below= .5cm of cat,align=center] (cls){\includegraphics[width=.4\textwidth]{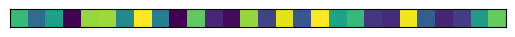} \\ Class Token};

\node[rect,below = .75cm of final,fill=orange] (prob){Probabilities};
\node[above left = .15cm of prob.north,thick]{Sigmoid} ;
\node[below = 0cm of enc, align=center]{ x \\ N};
\path[->,thick]
(final) edge (prob) (pos) edge (plus) (cls) edge (cat);

\end{tikzpicture}
}
\caption{The architecture for the (\textbf{a}) column-wise pooling and (\textbf{b}) the class-token models. For clarity, we use an MNIST image \cite{MNIST_DATA} to demonstrate the process. The hybrid and the classical model differ by the architecture of their encoder layers (see Figures~\ref{fig:encoder_classic} and \ref{fig:encoder_quantum}).}
\label{fig:column_architecture}
\end{figure}

Both the benchmark and hybrid models utilize the same architectures except for the type of encoder layers. These architectures are shown in Figure~\ref{fig:column_architecture}. As can be seen in the figure, there will be two main variants of the architecture: (a) column-pooling variant and (b) class token variant.

As the encoder layer is the main component of both the classical and the hybrid models, they will be discussed in more detail in Sections \ref{subsection:classic} and \ref{subsection:hybrid}, respectively. The rest of the architecture is discussed here.

First, we start by dividing our input picture into $n_t$ patches of equal area, which are then flattened to obtain $n_t$ vectors with length $d_i$. The resulting vectors are afterward concatenated to obtain a $n_t \times d_i$ matrix for each image sample. This matrix is passed through a linear layer with a bias (called "Linear Embedding" in the figure) to change the number of columns from $d_i$ to a desirable number (token dimension, referred to as $d_t$). 

If the model is a class token variant, a trainable vector of length $d_t$ is concatenated as the first row of the matrix at hand (module ``Concat'' in Figure~\ref{fig:column_architecture}b). After that, a non-trainable vector is added to each row (called the positional embedding vector). Then the result is fed to a series of encoder layers where each subsequent encoder layer uses its predecessor's output as its input.

If the model is a class token variant, the first row of the output matrix of the final encoder layer is fed into the classifying layer to obtain the classification probabilities (``Extract Class Token'' layer in Figure~\ref{fig:column_architecture}b). Otherwise, a column-pooling method (take the mean of all the rows or take the maximum value for each column) is used to reduce the output matrix into a vector, then this vector is fed into the classifying layer to obtain the classification probabilities (``Column-wise Pooling'' layer in Figure~\ref{fig:column_architecture}a).

\subsection{The Classical Encoder Layer}
\label{subsection:classic}

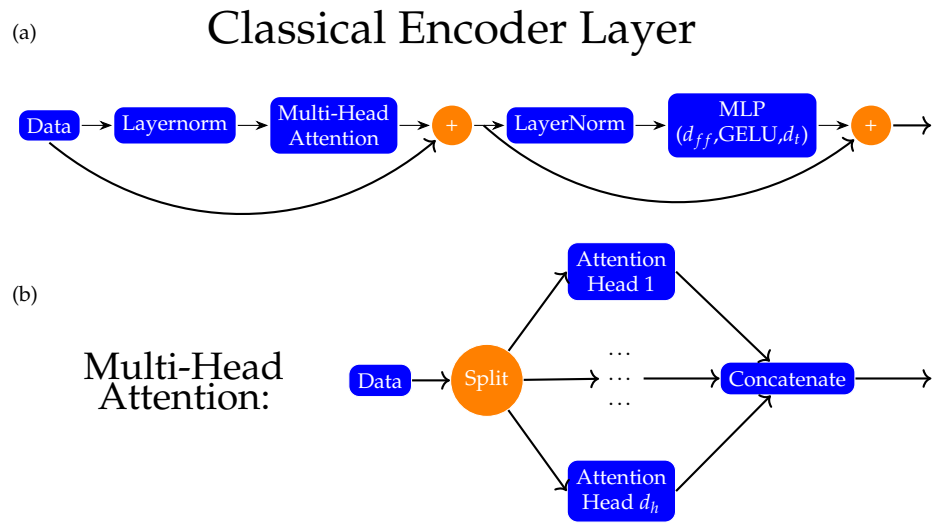
\begin{figure}[H]

\begin{tikzpicture}[
    node distance = 2mm and 4mm,
      start chain = going right,
  circ/.style = {shape=circle,  draw,color=white,fill=orange},
  rect/.style = {rounded corners,draw, align=center,color=white,fill=blue}]
\tikzstyle{every node}=[font=\footnotesize]
    \begin{scope}[every node/.append style={on chain,thick, join=by -Stealth}]

\node (n1) [rect]  {Data};
\node (n2) [rect]  {Layernorm};
\node (n3) [rect]  {Multi-Head \\Attention};
\node (n4) [circ]  {+};
\node (n5) [rect]  {LayerNorm};
\node (n6) [rect]  {MLP\\($d_{ff}$,GELU,$d_t$)};
\node (n7) [circ]  {+};

    \end{scope}
\node [right = .0cm of n4](x){};
\node [right = .5cm of n7](y){};
\node [above = .5cm of n4](title) {\LARGE Classical Encoder Layer};

  \node [left = 2cm of title](a) {(a)};

\path[->,thick]
(n1.south) edge[bend right=40] (n4) (x.center) edge[bend right=40] (n7)  (n7) edge (y);


\end{tikzpicture}

\begin{tikzpicture}[
  circ/.style = {shape=circle,  draw,color=white,fill=orange},
  rect/.style = {rounded corners,draw, align=center,color=white,fill=blue}]
\tikzstyle{every node}=[font=\footnotesize]
\node [align=center](n0) {\Large Multi-Head\\ \Large Attention:};
\node (n1) [left = 1cm of n0,rect] at (4,0,0)  {Data};
\node (n2) [right = .5cm of n1,circ] {Split};

\node (n3)  [above right = 1cm of n2,rect]   {Attention\\ Head 1};
\node (n4) [below right =1cm of n2,rect] {Attention\\ Head $d_h$};
\node (n5) [below =.55cm of n3,align=center] {\dots \\  \dots  \\ \dots};
\node (n6) [right =1cm of n5,align=center,rect] {Concatenate};
\node (n7) [right =1cm of n6] {};
\node [above left = .5cm of n0](b) {(b)};

\path[->,thick]
(n1) edge (n2) (n2) edge (n3.west) (n2) edge (n4.west) (n2) edge (n5);
\path[->,thick]
(n3.east) edge (n6) (n4.east) edge (n6) (n5.east) edge (n6) (n6) edge (n7) ;

\end{tikzpicture}

\caption{The classical encoder layer (\textbf{a}) and multi-head attention (\textbf{b}) architecture for the benchmark~models.
}
\label{fig:encoder_classic}
\end{figure}

The structure of the classical encoder layer can be seen in Figure~\ref{fig:encoder_classic}a. 
First, we start by standardizing the input data to have zero mean and a standard deviation of one. Afterward, the normalized data are fed to the multi-head attention (discussed in the next paragraph) and the output is summed with the unnormalized data. Then, the modified output is again normalized to have zero mean and a standard deviation of one. These normalized modified data are then fed into a multilayer perceptron of two layers with hidden layer size $d_{ff}$ and the result is summed up with the modified data to obtain the final result.
\vspace{-6pt}

The multi-head attention works by separating our input matrix into $n_h$ many $n_t \times d_h$ matrices by splitting them through their columns. Afterward, the split matrices are fed to the attention heads described in Equations~(\ref{eq:attentionhead}) and (\ref{eq:3}). Finally, the outputs of the attention heads are concatenated to obtain an $n_t \times d_t$ matrix, which has the same size as our input matrix. Each attention head is defined as
\begin{align}
\text{Attention Head }(x_i;W_K^{(i)},W_Q^{(i)},W_V^{(i)})&= \text{SoftMax} \left( \frac{(x_iW_K^{(i)})(x_i W_Q^{(i)})^T}{\sqrt{d_h}} \right) (x_iW_V^{(i)})\nonumber
\\
W_K^{(i)} \in R^{(d_h\times d_h)},W_Q^{(i)} &\in \mathbb{R}^{(d_h\times d_h)},W_V^{(i)} \in \mathbb{R}^{(d_h\times d_h)} \ d_h \equiv d_t/n_h;
\label{eq:attentionhead}
\end{align}
where 
\begin{equation}
X=\begin{bmatrix}
x_1 & x_2 & ... & x_{n_h} 
\end{bmatrix} \in \mathbb{R}^{(n_t \times d_t)}, \quad x_i\in \mathbb{R}^{(n_t \times d_h)}
\label{eq:3}
\end{equation}
is the input matrix.

\subsection{Hybrid Encoder Layer}
\label{subsection:hybrid}

\begin{figure}[H]

\begin{tikzpicture}[
    node distance = 2mm and 4mm,
      start chain = going right,
  circ/.style = {shape=circle,  draw,color=white,fill=orange},
  rect/.style = {rounded corners,draw, align=center,color=white,fill=blue}]
\tikzstyle{every node}=[font=\footnotesize]
    \begin{scope}[every node/.append style={on chain,thick, join=by -Stealth}]

\node (n1) [rect]  {Data};
\node (n2) [rect]  {Layernorm};
\node (n3) [rect]  {Hybrid \\ Multi-Head \\Attention};
\node (n6) [rect]  {MLP\\($d_{ff}$,GELU,$d_t$)};
\node (n7) [circ]  {+};

    \end{scope}
\node [right = .05cm of n4](x){};
\node [right = .5cm of n7](y){};
\node [left= 2cm of title](a) {(a)};
  
\node [above=.5cm of n6,align=center](title) {\LARGE Hybrid Encoder Layer};

\path[->,thick]
(n1.south) edge[bend right=40] (n7) (n7) edge (y);


\end{tikzpicture}

\begin{tikzpicture}[
  circ/.style = {shape=circle,  draw,color=white,fill=orange},
  rect/.style = {rounded corners,draw, align=center,color=white,fill=blue}]
\tikzstyle{every node}=[font=\footnotesize]
\node [align=center](n0) {\Large Multi-Head\\ \Large Attention:};
\node (n1) [left = 1cm of n0,rect] at (4,0,0)  {Data};
\node (n2) [right = .5cm of n1,circ] {Split};

\node (n3)  [above right = 1cm of n2,rect]   {Hybrid\\ Attention\\ Head 1};
\node (n4) [below right =1cm of n2,rect] {Hybrid\\ Attention\\ Head $d_h$};
\node (n5) [below =.55cm of n3,align=center] {\dots \\  \dots  \\ \dots};
\node (n6) [right =1cm of n5,align=center,rect] {Concatenate};
\node (n7) [right =1cm of n6] {};
\node [above left = .5cm of n0](b) {(b)};

\path[->,thick]
(n1) edge (n2) (n2) edge (n3.west) (n2) edge (n4.west) (n2) edge (n5);
\path[->,thick]
(n3.east) edge (n6) (n4.east) edge (n6) (n5.east) edge (n6) (n6) edge (n7) ;

\end{tikzpicture}

\caption{The hybrid encoder layer architecture (\textbf{a}) and multi-head attention (\textbf{b}) architecture for the hybrid models.
}
\label{fig:encoder_quantum}
\end{figure}
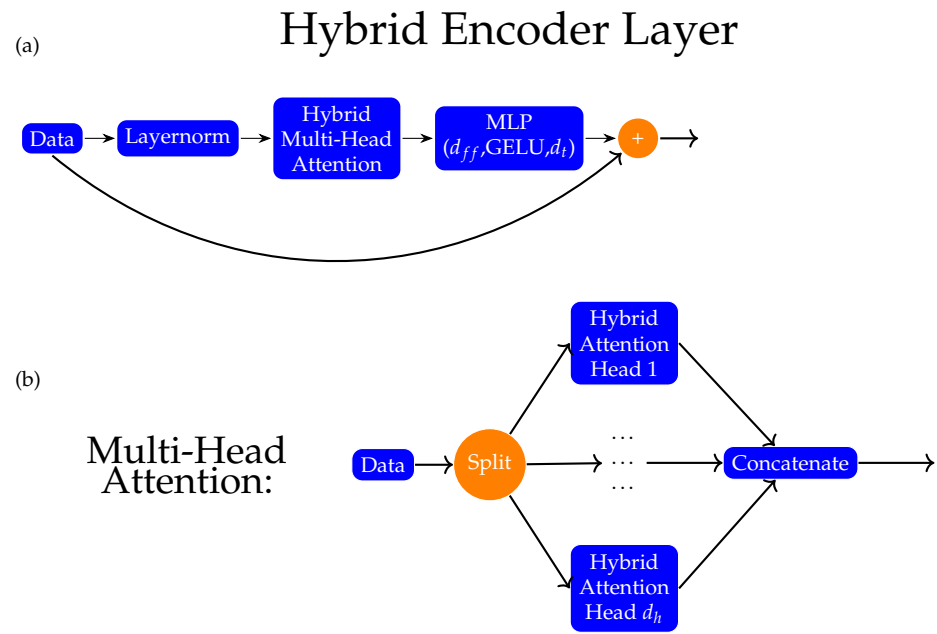

The structure of the hybrid encoder layer can be seen in Figure~\ref{fig:encoder_quantum}a. 
Firstly, we start by standardizing the input data to have zero mean and standard deviation of one. Afterward, the normalized data are fed to the hybrid multi-head attention layer (discussed in the next paragraph). Then, the output is fed into a multilayer perceptron of two layers with hidden layer size $d_{ff}$, and the result is summed up with the unnormalized data to obtain the final~result.

The hybrid multi-head attention works by separating our input matrix into $n_h$ many $n_t \times d_h$ matrices by splitting them through their columns. Afterward, the split matrices are fed to the hybrid attention heads (which are described in the bulleted procedure below). Finally, the outputs of the attention heads are concatenated to obtain an $n_t \times d_t$ matrix, which has the same size as our input matrix.

\begin{figure}[H]
 
\includegraphics[width=\textwidth]{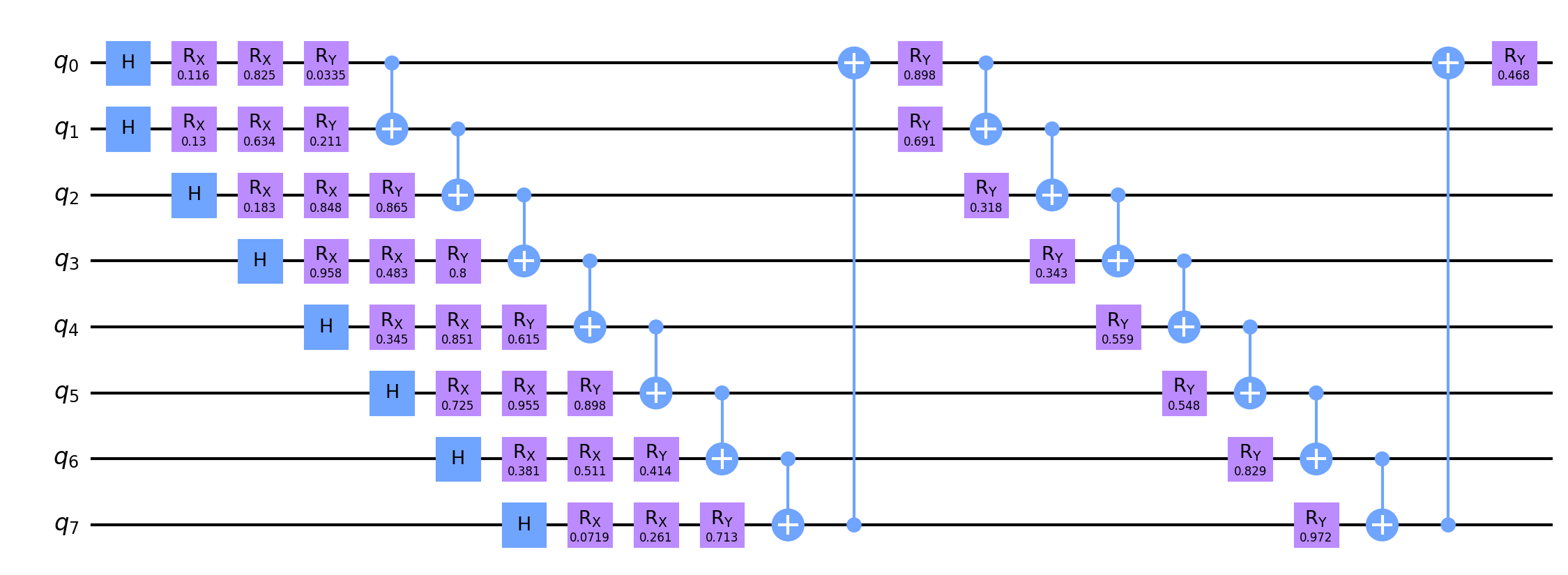}
\caption{Key and Query circuit for the $d_h=8$ case. The first two rows of circuits load the data to the circuit ($\hat U(x_i)$ operator), while the rest are the parts of the trainable ansatz. Therefore, the total number of parameters for each circuit is equal to $3d_h+1.$}
\label{fig:key_query}
\end{figure}

\begin{figure}[H]
 \hspace{-6mm}
\includegraphics[width=\textwidth]{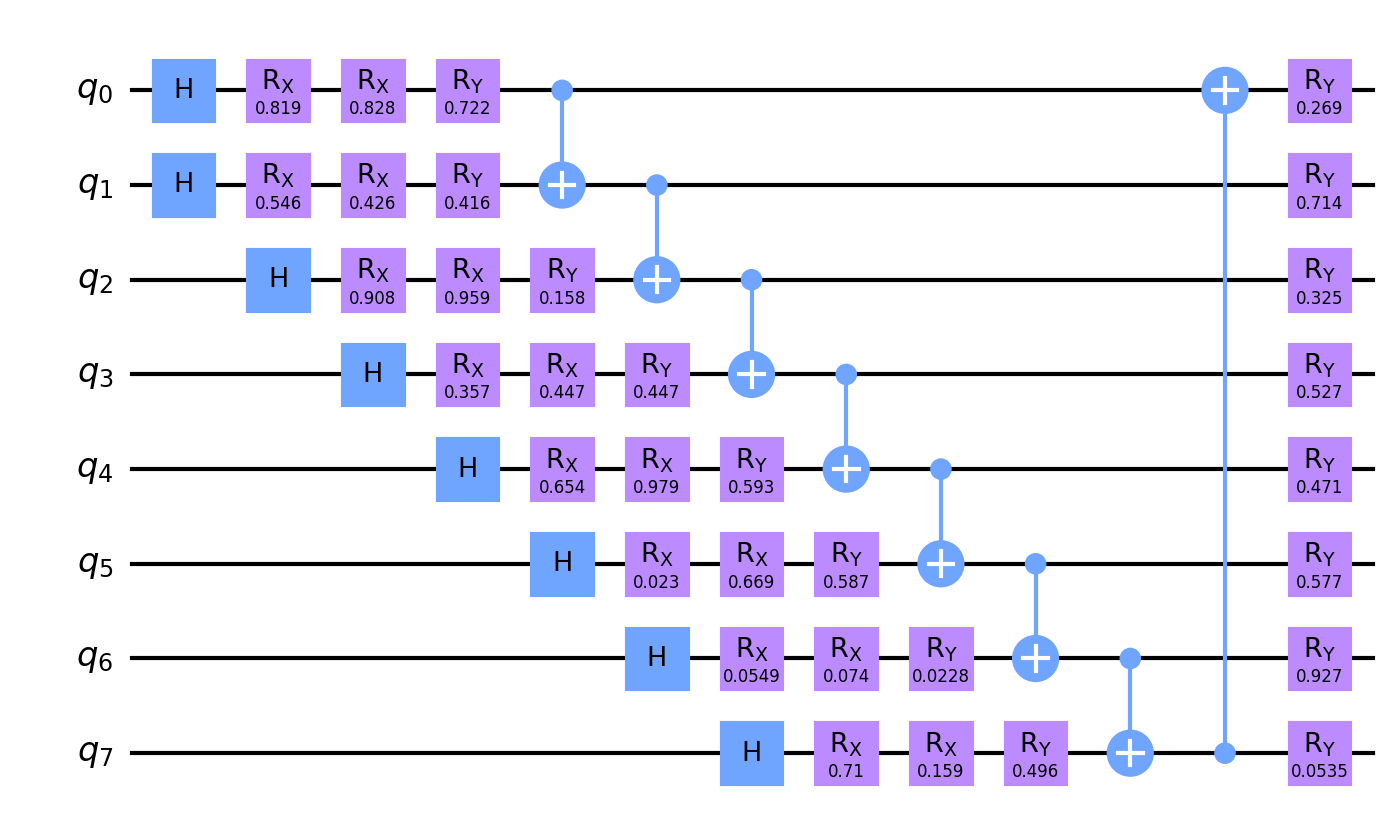}
\caption{The value circuit used for the $d_h=8$ case. The first two rows of circuits load the data to the circuit ($\hat U(x_i)$ operator), while the rest are the parts of the trainable ansatz. Therefore, the total number of trainable parameters for each circuit is equal to $3d_h$.}
\label{fig:value_circuit}
\end{figure}

The hybrid attention heads we used are almost identical to the architecture implemented in \cite{QSAN}, ``Quantum Self-Attention Neural Networks for Text Classification'' by Li et al. In order to replace the self-attention mechanism of a classical vision transformer in Equation~(\ref{eq:attentionhead}),
we use the following procedure: 
\begin{itemize}
\item Define $x_i$ as the $i$th row of the input matrix X.  
\item Define the data loader operator $\hat U(x_i)$ as 
\begin{equation}
\ket{x_i} \equiv \hat U(x_i)|0>^{(d_h)}=\bigotimes_{j=1}^{d_h}\hat R_x(x_{ij}) \hat H \ket{0},
\end{equation}
where $\hat H$ is the Hadamard gate and $\hat R_x$ is the parameterized rotation around the x-axis.

\item Apply the key circuit (data loader + key operator $\hat K(\theta_K)$) for each $x_i$ and obtain the column vector K (see Figure~\ref{fig:key_query}).
\begin{equation}
K_i = \bra{x_i}\hat K^\dag(\theta_K) \hat Z_0 \hat K(\theta_K) \ket{x_i},\quad 1\leq i\leq d_t,
\end{equation}
where $\hat Z_i$ is a spin measurement of the $i$th qubit on the z direction.
\item Apply the query circuit (data loader $\hat U(x_i)$ + query operator $\hat Q(\theta_Q)$) for each $x_i$ and obtain the column vector Q (see Figure~\ref{fig:key_query}).
\begin{equation}
Q_i = \bra{x_i}\hat Q^\dag(\theta_Q) \hat Z_0 \hat Q(\theta_Q) \ket{x_i}, \quad 1\leq i\leq d_t.
\end{equation}
\item Obtain the so-called attention matrix using the key and the query vectors using the following expression
\begin{equation}
A_{ij} = -(Q_i-K_j)^2; \quad 1\leq i\leq d_t, 1\leq j\leq d_t.
\end{equation}
\item Apply the value circuit (data loader + value operator $\hat V(\theta_V)$) to each row of the image and measure each qubit separately to obtain the value matrix. (See Figure~\ref{fig:value_circuit})
\begin{equation}
V_{ij} = \bra{x_i}\hat V^\dag(\theta_V) \hat Z_j \hat V(\theta_V) \ket{x_i}, \ket{x_i} = \hat U(x_i)|0_n> ; \quad 1\leq i\leq d_t, 1\leq j\leq d_h.
\end{equation}
\item Define the self-attention operation as,
\begin{equation}
\text{Hybrid Attention Head: } \text{SoftMax} \left( \frac{A}{\sqrt{d_h}} \right) V.
\end{equation}

\end{itemize}

\vspace{-16pt}

\section{Hyper-Parameters}
\label{section:hyper}

The number of parameters is a function of the hyper-parameters for both the classical and the hybrid models. However, these functions are different.
Both models share the same linear embedding and classifying layer. The linear embedding layer contains $(d_i+1)d_t$ many parameters and the classifying layer contains $32d_t+65$ parameters.

For each classical encoder layer, we have $n_h$ attention heads which all contain $3d_h^2$ parameters from the Q, K, and V layers, respectively. In addition, the MLP layer inside each encoder layer contains $2 d_{ff}d_t+d_{ff}+d_t$ parameters.
Overall, each classical vision transformer has
$d_t(33+d_i)+n_l(2d_{ff}d_t+d_{ff}+d_t+3n_hd_h^2)$ parameters except for the class token variation which has extra $d_t$ parameters.

For each hybrid encoder layer, we have $n_h$ attention heads which all contain $9d_h+2$ parameters from the Q, K, and V layers, respectively. {Similar to the classical model, each encoder layer MLP} contains $2 d_{ff}d_t+d_{ff}+d_t$ parameters.
Overall, each hybrid vision transformer has
$d_t(33+d_i)+n_l(2d_{ff}d_t+d_{ff}+d_t+n_h(9d_h+2))$ parameters except for the class token variation which has extra $d_t$ parameters.

Therefore, assuming they have the same hyper-parameters, the difference between the number of parameters for the classical and hybrid models is $n_l(d_t(3d_h-9)-2n_h)$.

Our purpose was to investigate whether our architecture might perform similarly to a classical vision transformer where the number of parameters are close to each other. In order to use a similar number of parameters, we picked a region of hyperparameters such that this difference is rather minimal. For all models, the following parameters were used:
\begin{itemize}
\item $n_l=5$
\item $d_t=16$
\item $n_t=16$
\item $n_h=4$
\item $d_h=\frac{d_i}{d_h} = 4$
\item $d_{ff}=16.$
\end{itemize}
\newpage
Therefore, for our experiment the number of parameters for the classical models (4785~to 4801) is slightly more than the quantum models (4585 to 4601).

\section{Training Process}
\label{section:training}
\begin{figure}[H]
\centering
(a)

\includegraphics[width=.34\textwidth]{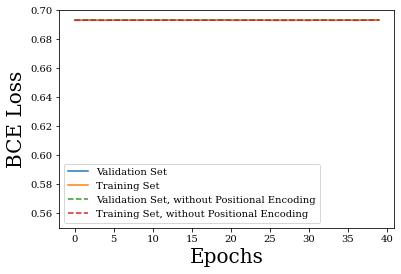}
\includegraphics[width=.32\textwidth]{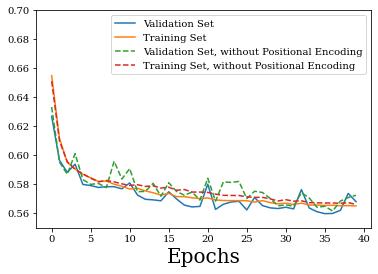}
\includegraphics[width=.32\textwidth]{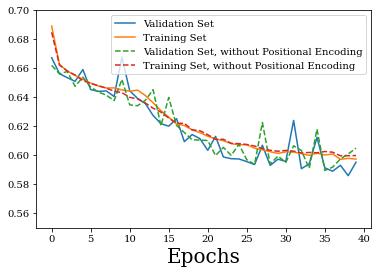}

(b)

\includegraphics[width=.34\textwidth]{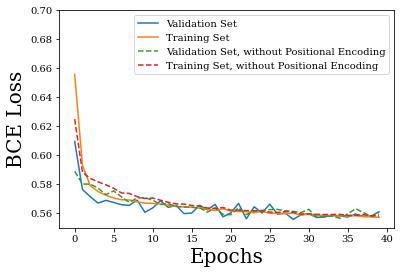}
\includegraphics[width=.32\textwidth]{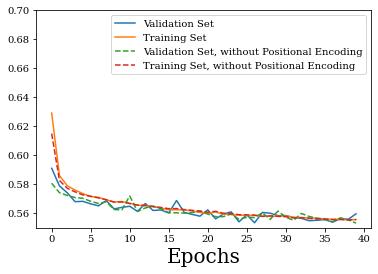}
\includegraphics[width=.32\textwidth]{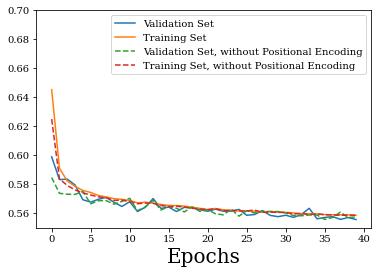}
\caption{BCE loss on the validation and training set during training for the (\textbf{a}) quantum and (\textbf{b})~classical models. From left to right, each column corresponds to a different model variant: class token (left column), column max (middle column) and column mean variant (right column). For each plot, the blue (orange) line corresponds to the validation (training) set loss for the model with positional encoding, whereas the dashed green (red) line corresponds to the validation (training) set loss for the model without positional encoding layer.}
\label{fig:tr_comp_cls}
\end{figure}

\begin{figure}[H]
\centering

(a)

\includegraphics[width=.34\textwidth]{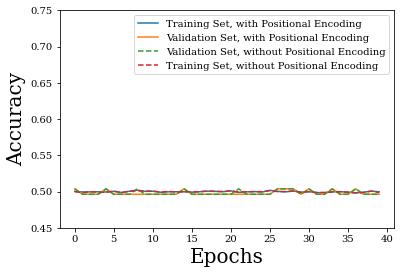}
\includegraphics[width=.32\textwidth]{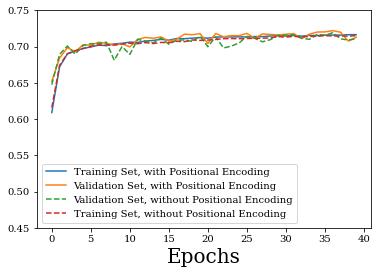}
\includegraphics[width=.32\textwidth]{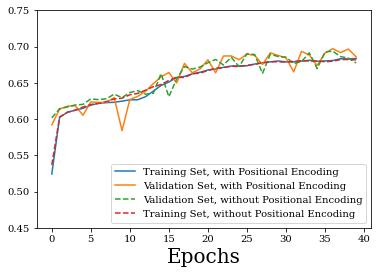}

(b)

\includegraphics[width=.34\textwidth]{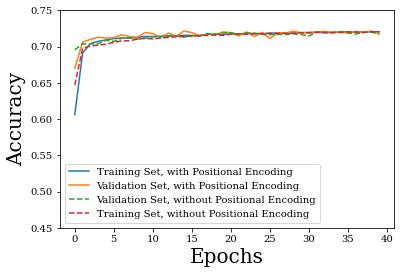}
\includegraphics[width=.32\textwidth]{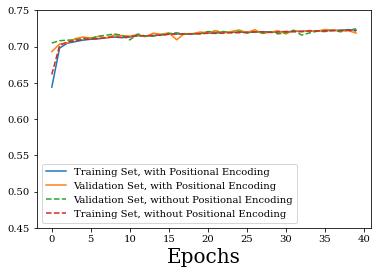}
\includegraphics[width=.32\textwidth]{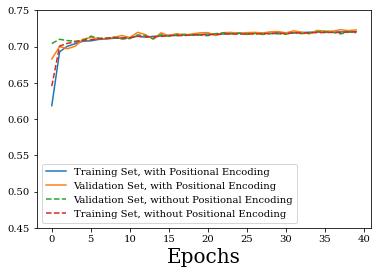}
\caption{The same as Figure~\ref{fig:tr_comp_cls}, but for the accuracy on the validation and training set during training for the (\textbf{a}) quantum and (\textbf{b}) classical models. 
}
\label{fig:tr_comp_qml}
\end{figure}

All the classical parts of the models were implemented in PyTorch \cite{torch}. The quantum circuit simulations were conducted using TensorCircuit with the JAX backend \cite{jax,tensorcircuit}. {We explored a few different hyperparameter settings before settling on the following.}
Each model was trained for 40 epochs, {which was typically sufficient to ensure convergence, see Figures~\ref{fig:tr_comp_cls} and \ref{fig:tr_comp_qml}}. The criteria for the selection of the best model iteration was the accuracy of the validation data. The optimizer used was the ADAM optimizer with learning rate $\lambda = 5 \times10^{-3}$ \cite{adam}. All models were trained on GPUs {and the typical training times were on the order of 10 min (5 h) for the classical (quantum) models}. The batch size was 512 for all models as well. The loss function utilized was the binary cross entropy.
The code used to create and train the models can be found at the following GitHub repository:
\url{https://github.com/EyupBunlu/QViT_HEP_ML4Sci} (accessed on 7 March 2024).

\vspace{-6pt}

\vspace{-6pt}

\section{Results}
\label{section:results}

The training loss and the accuracy of the validation and training data are plotted in Figures \ref{fig:tr_comp_cls} and \ref{fig:tr_comp_qml}, respectively. In addition, the models were compared on several metrics such as the accuracy, binary cross-entropy loss, and AUC (area under the ROC curve) on the test data. This
comparison is shown in Table \ref{table:t}.
\begin{table}[H]
\caption{Comparison table for the models. The accuracy, the BCE loss and the AUC score were calculated on the test data. For each entry, the first number corresponds to the classical model, whereas the second one corresponds to the hybrid model. For each variant and metric, the best value is shown in bold.}
\label{table:t}
\begin{tabularx}{\textwidth}{lXXXXXX}
\toprule
\textbf{Model} & \textbf{Positional} \newline {\textbf{Embedding}} & \textbf{Accuracy} \newline \textbf{(Cls/Hybrid)}& \textbf{BCE Loss} \newline \textbf{(Cls/Hybrid)}& \textbf{AUC Score} \newline \textbf{(Cls/Hybrid)}& \textbf{Trainable} \newline \textbf{Parameters} \newline \textbf{(Cls/Hybrid)} \\ \midrule
With Class Token & Yes & {\bf 0.717}/0.502 & {\bf 0.564}/0.6931 & {\bf 0.780}/0.501 & 4801/4601 \\ \midrule
With Class Token & No  & {\bf 0.720}/0.502 & {\bf 0.561}/0.6931 & {\bf 0.783}/0.500 & 4801/4601\\ \midrule
Column Max (CMX)      & Yes & {\bf 0.718}/{\bf 0.718} & {\bf 0.562}/0.565 & {\bf 0.783}/0.779 & 4785/4585\\ \midrule
Column Max  (CMX)     & No  & {\bf 0.722}/0.718 & {\bf 0.557}/0.565 & {\bf 0.786}/0.779 & 4785/4585\\ \midrule
Column Mean  (CMN)    & Yes & {\bf 0.720}/0.696 & {\bf 0.559}/0.592 & {\bf 0.784}/0.751 & 4785/4585\\ \midrule
Column Mean  (CMN)    & No  & {\bf 0.720}/0.692 & {\bf 0.560}/0.595 & {\bf 0.783}/0.748 & {4785/4585}\\ \bottomrule
\end{tabularx}

\end{table}

\section{Discussion}
\label{section:discussion}

As seen in Table~\ref{table:t}, the positional encoding has no significant effect on the performance metrics. {In retrospect, this is not that surprising, since the position information is already used in the linear embedding in Figure~\ref{fig:column_architecture} (we thank the anonymous referee for this clarification.)}. 
We note that the CMX variant (either with or without positional encoding) performs similarly to the corresponding classical model. This suggests that a quantum advantage could be achieved when extrapolating to higher-dimensional problems and datasets since the quantum models scale better with dimensionality.


On the other hand, Table~\ref{table:t} shows that hybrid CMN variants are inferior to their hybrid CMX counterparts for all metrics. This might be due to the fact that taking the mean forces each element of the output matrix of the final encoder layer to be relevant, unlike the CMX variant, where the maximum values are chosen. This could explain the larger number of epochs required to converge in the case of the hybrid CMN (see Figures~\ref{fig:tr_comp_cls} and~\ref{fig:tr_comp_qml}). It is also possible that the hybrid model lacks the expressiveness required to encode enough meaningful information to the column means.

Somewhat surprisingly, the training plots of the hybrid class token variants (upper left panels in Figures~\ref{fig:tr_comp_cls} and~\ref{fig:tr_comp_qml}) show that the hybrid class token variants did not converge during our numerical experiments. The reason behind this behavior is currently unknown and is being investigated.

\section{Outlook}
\label{sec:outlook}

Quantum machine learning is a relatively new field. In this work, we explored a few of the many possible ways that it could be used to perform different computational tasks as an alternative to classical machine learning techniques. As the current hardware for quantum computers improves further, it is important to explore more ways in which this hardware could be utilized.

Our study raises several questions that warrant future investigations.
First, we observe that the hybrid CMX models perform similarly to the classical vision transformer models that we used for benchmarking. It is fair to ask if this similarity is due to the comparable number of trainable parameters or the result of an identical choice of hyper-parameter values. If it is the latter, we can extrapolate and conclude that as the size of the data grows, hybrid models will still perform as well as the classical models while having a significantly fewer number of parameters. 

It is fair to say that both the classical and hybrid models perform similarly at this scale. However, the hybrid model discussed in this work is mostly classical, except for the attention heads. The next step in our research is to investigate the effect of increasing the fraction of quantum elements of the model. For instance, the conversion of feed-forward layers into quantum circuits such as the value circuit might lead to an even bigger advantage in the number of trainable parameters between the classical and hybrid models.

Although the observed limitations in the class token and column mean variants might appear disappointing at first glance, they are also important findings of this work. It is worth investigating whether this is due to the nature of the dataset or a sign of a fundamental limitation in the method. 

\vspace{6pt} 



\authorcontributions{Conceptualization, E.B.U.; methodology, M.C.C., G.R.D., Z.D., R.T.F., S.G., D.J., K.K., T.M., K.T.M., K.M. and E.B.U.; software, E.B.U.; validation, M.C.C., G.R.D., Z.D., R.T.F., T.M. and E.B.U.; formal analysis, E.B.U.; investigation, M.C.C., G.R.D., Z.D., R.T.F., T.M. and E.B.U.; resources, E.B.U., K.T.M. and K.M.; data curation, G.R.D., S.G. and T.M.; writing---original draft preparation, E.B.U.; writing---review and editing, S.G., D.J., K.K., K.T.M. and K.M.; visualization, E.B.U.;
supervision, S.G., D.J., K.K., K.T.M. and K.M.; project administration, S.G., D.J., K.K., K.T.M. and K.M.; funding acquisition, S.G. All authors have read and agreed to the published version of the~manuscript. }

\funding{This research used resources of the National Energy Research Scientific Computing Center, a DOE Office of Science User Facility supported by the Office of Science of the U.S. Department of Energy under Contract No. DE-AC02-05CH11231 using NERSC award NERSC DDR-ERCAP0025759. S.G. is supported in part by the U.S. Department of Energy (DOE) under Award No. DE-SC0012447. K.M. is supported in part by the U.S. DOE award number DE-SC0022148. K.K. is supported in part by US DOE DE-SC0024407. Z.D. is supported in part by College of Liberal Arts and Sciences Research Fund at the University of Kansas. Z.D., R.T.F., E.B.U., M.C.C., G.R.D. and T.M. were participants in the 2023 Google Summer of Code.}

\institutionalreview{Not applicable.}


\dataavailability{The dataset used in this analysis is described in \cite{Andrews:2018gew} and is available at\linebreak \url{https://cernbox.cern.ch/index.php/s/FbXw3V4XNyYB3oA }  (accessed on 12 March 2024) and \url{https://cernbox.cern.ch/index.php/s/AtBT8y4MiQYFcgc} (accessed on 12 March 2024). The code used to create and train the models can be found at \url{https://github.com/EyupBunlu/QViT_HEP_ML4Sci} (accessed on 12 March 2024).}


\conflictsofinterest{The authors declare no conflicts of interest. The funders had no role in the design of the study; in the collection, analyses, or interpretation of data; in the writing of the manuscript; or in the decision to publish the results.} 



\abbreviations{Abbreviations}{
The following abbreviations are used in this manuscript:\\

\noindent 
\begin{tabular}{@{}ll}
AUC & Area Under the Curve \\
BCE & Binary Cross Entropy\\
CMN & Column Mean\\
CMS & Compact Muon Solenoid (experiment)\\
CMX & Column Max\\
ECAL & Electromagnetic Calorimeter\\
GPU & Graphics processing unit\\
LHC & Large Hadron Collider\\
MHA & Multi-Head Attention \\
MLP & Multi-Layer Perceptron \\
MNIST & Modified National Institute of Standards and Technology database\\
ROC & Receiver Operating Characteristic
\end{tabular}
}

\reftitle{References}



\begin{adjustwidth}{-\extralength}{0cm}

\PublishersNote{}
\end{adjustwidth}

\end{document}